# Investigation to implicate data on clouds

Nidhi Bansal
nidhi18jul@gmail.com

B.Tech (CSE) from Shobhit Institute of Engineering & Technology (SIET) Meerut

**Abstract**

Cloud computing can and does mean different things to different people. The common characteristics most shares are on-demand secure access to metered services from nearly anywhere and dislocation of data from inside to outside the organization. Vision of cloud computing as a new IT procurement model. The system lifecycle, risks that are identified must be carefully balanced against the security and privacy controls available and the expected benefits from their utilization. Too many controls can be inefficient and ineffective, if the benefits outweigh the costs and associated risks.

In this micro research, we characterize the problems related to security challenges.

**Related work**

Cloud providers are generally not aware of a specific organization's security and privacy needs. Adjustments to the cloud computing environment may be warranted to meet an organization's requirements. Organizations should require that any selected public cloud computing solution is configured, deployed, and managed to meet their security, privacy, and other requirements.

**Proposed Work**

Want to show the different eye direction to understand this cloud services to provide security for proper channel on data processing over clouds.

**Introduction**

Cloud computing refers to data, processing power, or software stored on remote servers made accessible by the Internet as opposed to one's own computers. The term "the cloud" comes from computer network diagrams which, because the individual computers that formed its components were too numerous to show individually, depicted the Internet as a vast cloud at the top of the network chain. One of the key features of cloud computing is that the end users does not own the technology they are using. All the hardware and software is owned by a cloud computing service, while the user simply rents time or space. Several cloud computing applications, such web email, wiki applications, and online tax preparation, have become common experiences for the average Internet user.

For users, cloud computing arrangements can bring about major cost reductions and efficiencies. For example, in a cloud computing arrangement the end user does not have to pay large up front capital costs for hardware or for that hardware's continued maintenance. If the user needs temporary additional space, he can simply tell the cloud service provider to up his quota for the time being, rather than purchase additional physical capacity which would only be needed for a short period and then left idle. This also means that computer resources as a whole are generally used more efficiently. Cloud Computing can be thought of as a way to make the world of computer resources seamlessly scalable.

Common thought, organizations should have security controls in place for cloud-based applications that are commensurate with or surpass those used if the applications were deployed in-house. Cloud computing is heavily dependent on the individual security of each of its many components, including those for self-service, quota management, and resource metering, plus the hypervisor, guest virtual machines, supporting middleware, deployed applications, and data storage. Many of the simplified interfaces and service abstractions belie the inherent complexity that affects security. Organizations should ensure to the extent practical that all of these elements are secure and that security is maintained based on sound security practices.

**What are the "security" concerns?**

**Let see the Example** of Google accounts, there are many accounts created in Google website to secure data and communicate for other party.

**Have you ever think about that where your Google account email data save or secured by whom?**

-You have many mails in your account, that mails are managed by server of TCS which is work for Google accounts.
Similarly:
Every site have own server to manage and secure your data.

**Still your account can be hacked by any one. Why?**
-this problem is caused by networking or person (not trustworthy).
-this problem is not done by the server (if any server crashed then no worry, from other replicated copy it'll be recovered by recovery manager)

**Now you are thinking about that on which server your data stored?**
We all know that our data is managed by the server but still we want to know that answer for this question.
I accept your query: on which server?
See: we don't know about the whole application or services that is introduced or created by the Google organization for our help.
Example: we can't remember the all port number for different sites; remember only that which ip we used.

**Cloud Services:** There are three basic types of cloud computing:

1. Software as a Service (**SaaS**) is the most common and widely known type of cloud computing. SaaS applications provide the function of software that would normally have been installed and run on the user's desktop. With SaaS, however, the application is stored on the cloud computing service provider's servers and run through the user's web browser over the Internet. Examples of SaaS include: Gmail, Google Apps, and Salesforce.
2. Platform as a Service (**PaaS**) cloud computing provides a place for developers to develop and publish new web applications stored on the servers of the PaaS provider. Customers use the Internet to access the platform and create applications using the PaaS provider's API, web portal, or gateway software. Examples of PaaS include: Saleforce's Force.com, Google App Engine, Mozilla Bespin, Zoho Creator.
3. Infrastructure as a Service (**IaaS**) seeks to obviate the need for customers to have their own data centers. IaaS providers sell customers access to web storage space, servers, and Internet connections. The IaaS provider owns and maintains the hardware and customers rent space according to their current needs. An example of Iaas is Amazon Web Services. IaaS is also known as utility computing.

**Transition mechanisms**

Transition mechanisms are needed to enable IPv6-only hosts to reach IPv4 services

- RFC 2185, Routing Aspects of IPv6 Transition
- RFC 2766, Network Address Translation — Protocol Translation NAT-PT, obsolete as explained in RFC 4966 Reasons to Move the Network Address Translator — Protocol Translator NAT-PT to Historic Status
- RFC 3053, IPv6 Tunnel Broker
- RFC 3056, 6to4. Connection of IPv6 Domains via IPv4 Clouds
- RFC 3142, An IPv6-to-IPv4 Transport Relay Translator
- RFC 4213, Basic Transition Mechanisms for IPv6 Hosts and Routers
- RFC 4380, Teredo: Tunneling IPv6 over UDP through Network Address Translations NATs
- RFC 4798, Connecting IPv6 Islands over IPv4 MPLS Using IPv6 Provider Edge Routers (6PE)
- RFC 5214, Intra-Site Automatic Tunnel Addressing Protocol ISATAP
- RFC 5569, IPv6 Rapid Deployment on IPv4 Infrastructures (6rd)
- RFC 5572, IPv6 Tunnel Broker with the Tunnel Setup Protocol (TSP)
- RFC 6180, Guidelines for Using IPv6 Transition Mechanisms during IPv6 Deployment
- RFC 6343, Advisory Guidelines for 6to4 Deployment

**NAT64** Network Address Translation/Protocol Translation (or simply **NAT-PT**) is a mechanism to allow IPv6 hosts to communicate with IPv4 servers

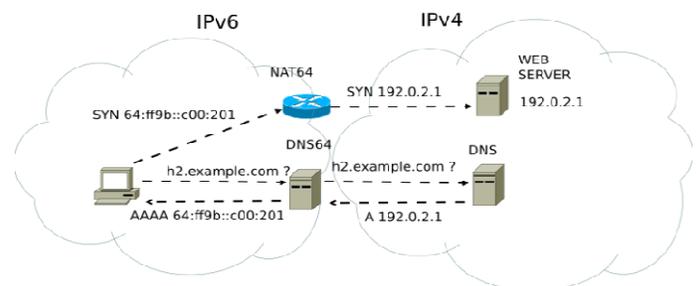

**We categorize the security concerns as:**
Traditional security
Availability
Third-party data control

**Security and Privacy Issues and Precautions Table:**

| Areas | Precautions |
|---|---|
| 1. Governance | Extend organizational practices pertaining to the policies, procedures, and standards used for application development and service provisioning in the cloud, as well as the design, implementation, testing, and monitoring of deployed or engaged services. Put in place audit mechanisms and tools to ensure organizational practices are followed throughout the system lifecycle. |
| 2. Compliance | Understand the various types of laws and regulations that impose security and privacy obligations on the organization and potentially impact cloud computing initiatives, particularly those involving data location, privacy and security controls, and electronic discovery requirements. Review and assess the cloud provider's offerings with respect to the organizational requirements to be met and ensure that the contract terms adequately meet the requirements. |
| 3. Trust | Incorporate mechanisms into the contract that allow visibility into the security and privacy controls and processes employed by the cloud provider, and their performance over time. Institute a risk management program that is flexible enough to adapt to the continuously Evolving and shifting risk landscape. |
| 4. Architecture | Understand the underlying technologies the cloud provider uses to provision services, including the implications of the technical controls involved on the security and privacy of the system, with respect to the full life cycle of the system and for all system components. |
| 5. Identity and Access Management | Ensure that adequate safeguards are in place to secure authentication, authorization, and other identity and access management functions. |
| 6. Software Isolation | Understand virtualization and other software isolation techniques that the cloud provider employs, and assess the risks involved. |
| 7. Data Protection | Evaluate the suitability of the cloud provider's data management solutions for the organizational data concerned. |
| 8. Availability | Ensure that during an intermediate or prolonged disruption or a serious disaster, critical operations can be immediately resumed and that all operations can be eventually reinstituted in a timely and organized manner |
| 9. Incident Response | Understand and negotiate the contract provisions and procedures for incident response required by the organization. |

**Common Threats:**
- Abuse and Nefarious Use of Cloud Computing
- Insecure Interfaces and APIs
- Malicious Insiders
- Shared Technology Issues
- Data Loss or Leakage
- Account or Service Hijacking

**A promising approach** to address this problem is based on Trusted Computing. Imagine a trusted monitor installed at the cloud server that can monitor or audit the operations of the cloud server. The trusted monitor can provide "proofs of compliance" to the data owner, stating that certain access policies have not been violated. To ensure integrity of the monitor, Trusted Computing also allows secure bootstrapping of this monitor to run beside (and securely isolated from) the operating system and applications. The monitor can enforce access control policies and perform monitoring/auditing tasks. To produce a "proof of compliance", the code of the monitor is signed, as well as a "statement of compliance" produced by the monitor. When the data owner receives this proof of compliance, it can verify that the correct monitor code is run, and that the cloud server has complied with access control policies.

A different approach to retaining control of data is to require the encryption of all cloud data. The problem is that encryption limits data use. In particular searching and indexing the data becomes problematic. For example, if data is stored in cleartext, one can efficiently search for a document by specifying a keyword. This is impossible to do with traditional, randomized encryption schemes. State-of-the-art cryptography may offer new tools to solve these problems. Cryptographers have recently invented versatile encryption schemes that allow operation and computation on the cipher text.

While in many cases more research is needed to make different cryptographic tools sufficiently practical for the cloud, we believe
they present the best opportunity for a clear differentiator for cloud computing since these protocols can enable cloud users to benefit from one another's data in a controlled manner. In particular, even encrypted data can enable anomaly detection that
is valuable from a business intelligence standpoint. For example, a cloud payroll service might provide, with the agreement of
participants, aggregate data about payroll execution time that allows users to identify inefficiencies in their own processes.
Taking the vision even further, if the cloud service provider is empowered with some ability to search the encrypted data, the
proliferation of cloud data can potentially enable better insider threat detection (e.g. by detecting user activities outside of the
norm) and better data loss prevention (DLP) (e.g. through detecting anomalous content).

## Conclusion

Cloud fears largely stem from the perceived loss of control of sensitive data. Current control measures do not adequately address cloud computing third-party data storage and processing needs. In our vision, we propose to extend control measures from the
enterprise into the cloud through the use of Trusted Computing and applied cryptographic techniques
Our vision also relates to likely problems and abuses arising from a greater reliance on cloud computing, and how to maintain
security in the face of such attacks. Namely, the new threats require new constructions to maintain and improve security.
Among these are tools to control and understand privacy leaks, perform authentication, and guarantee availability in the face of cloud denial-of-service attacks.

Accountability for security and privacy in public clouds remains with the organization. Federal agencies must ensure that any selected public cloud computing solution is configured, deployed, and managed to meet the security, privacy, and other requirements of the organization. Organizational data must be protected in a manner consistent with policies, whether in the organization's computing center or the cloud. The organization must ensure that security and privacy controls are implemented correctly and operate as intended.

**Result: See with different thinking and Trust your own Server.**

## Acknowledgements

The researchers wish to express their deepest gratitude and warmest appreciation to the following people, who, in any way have contributed and inspired the researchers to the overall success of the undertaking:

•To our friends, who have been unselfishly extending their efforts and understanding.

•To our parents who have always been very understanding and supportive both financially and emotionally

•and above all, to the Almighty God, who never cease in loving us and for the continued guidance and protection.